# High-Efficiency Isolator-Free Magnetron Power Combining Method Based on H-Plane Tee Coupling and Peer-to-Peer Locking

Shaoyue Wang, *Graduated Student Member, IEEE*, Xu Zhu, *Graduated Student Member, IEEE*,
Xiaojie Chen, Da He, *Graduated Student Member, IEEE*, Zhongqi He, *Member, IEEE*
Liping Yan, *Senior Member, IEEE,* and Changjun Liu, *Senior Member, IEEE*

*Abstract*—Magnetrons are widely used as high-performance microwave sources in microwave heating, microwave chemistry, and microwave power transmission due to their high efficiency, low cost, and compact size advantages. However, the output power of a single magnetron is limited by its resonant cavities, posing a physical constraint. High-efficiency coherent power combining based on the injection-locking technique effectively overcomes this limitation and meets the demand for higher output power. Nevertheless, using isolators such as circulators introduces significant insertion loss, and the injection signal sources and phase shifters increase the system's size, cost, and complexity in a conventional magnetron power combining system. A novel method is proposed to utilize the coupling between two ports of an H-plane tee to achieve peer-to-peer injection locking magnetrons. Meanwhile, an asymmetric phase compensation is realized using a section of waveguide to adjust the magnetrons' output characteristics. Theoretical and numerical analyses provided qualitative insight into the system's output behavior. Subsequently, an experimental system was developed for verification. In the experiments, the system achieved maximum microwave power combining efficiencies 90.2%, 93.6%, and 93.6% at electrical waveguide lengths corresponding to 90°, 135°, and 225°, with output powers of 1650 W, 1260 W, and 1610 W, respectively, without the use of any isolators or external injection sources. The experimental results show good agreement with numerical calculations. This method offers the advantages of low cost, compact size, and low loss, providing a new approach for developing high-performance magnetron power combining systems in the future.

## I. INTRODUCTION

MAGNETRONS are crucial components of microwave sources with the advantages of high efficiency, low cost, and compact size. They are widely used in various applications, such as microwave heating, microwave chemistry, microwave wireless power transmission (MWPT), and so on [1]-[5]. However, with the development of microwave technologies, the demand for large-power microwave sources is increasing in applications such as microwave heating and MWPT. Magnetrons' output power is limited by the size of their resonant cavities, which is directly related to their operating frequencies. The power capacity of a continuous wave magnetron is around 100 kW at 915 MHz, and 30 kW at 2.45 GHz [6]. Therefore, coherent power combining, as a technology that can overcome the magnetrons' physical power limit and meet the demand for higher microwave power, has attracted significant attention from researchers.

For high-efficiency coherent power combining, the involved signals typically need to maintain consistent frequency, phase, and amplitude [7]-[9]. Due to the significant frequency and phase jitter in free-running magnetrons, Injection-locking techniques, as an effective method for controlling magnetron output, have attracted significant research interest — both in understanding their characteristic [10][11] and in exploring their application for achieving high-efficiency magnetron power combining (MPC) systems.

Shahid *et al.* proposed a power combining scheme without phase shifter where a single signal generator synthesizing the injection signal for both magnetrons, achieving a combining efficiency of 95% [12]. The experimental Liu *et al.* developed a dual-way coherent power combining system utilizing external injection and real-time phase control, achieving a combining efficiency of up to 96.6% and a continuous wave output power of 26.3 kW [13]. They later proposed a four-way, 20-kW injection-locked magnetron coherent power combining system, achieving a maximum output power of 60.6 kW with a combining efficiency of 91.5% [14]. Similarly, Huang *et al.* designed a power combining system using a five-port power combiner, reaching a combining efficiency of 97.7% with four externally injection-locked magnetrons [15].

Although injection-locking-based phase control enables high combining efficiency, the insertion loss of isolation components such as circulators is significant, reducing overall system efficiency while increasing both cost and size. Consequently, avoiding using circulators or finding suitable alternatives has become a key research focus. Terado was the first to replace the circulator and power combiner with a 3-dB coupler, achieving a long-pulse power combining with an efficiency of 92% [16]. Yuan *et al.* established an MPC

This work was supported in part by the National Natural Science Foundation of China (NSFC) under Grant U22A2015.
S. Wang, X. Zhu, D. He, Z. He, L. Yan and C. Liu are with the College of Electronics and Information Engineering, Sichuan University, Chengdu 610064, China (Corresponding author: *Changjun Liu*. e-mail: cjliu@ieee.org).
X. Chen is with the High People's Court of Guangxi Autonomous Region, Nanning 530028, China

system based on master–slave injection locking, which operated without signal sources or real-time phase control, and finally achieved a combining efficiency exceeding 96% [17]. Chen *et al.* replaced traditional H-plane tee or E-plane tee components with a waveguide magic tee, eliminating the need for circulators and significantly reducing system losses, achieving an efficiency of 94.5% [18]. Similarly, Shahid *et al.* realized four-way coherent power combining without circulators by employing three magic-tee combiners, achieving an efficiency of 93% [19]. Zhang *et al.* proposed a power combining system based on quasi-locked magnetrons using master-slave injection. By incorporating a waveguide phase shifter for phase adjustment, they constructed a system with only one circulator, achieving a combining efficiency of 90% [20].

Peer-to-peer injection refers to coupling the outputs of two magnetrons through a common transmission path, enabling mutual frequency and phase locking. Pengvanich conducted a theoretical analysis of the general model of peer-to-peer injection-locking [21], which was later experimentally verified by Cruz [22]. In recent years, peer-to-peer injection-locked magnetrons have garnered increasing research interest. Liu *et al.* realized peer-to-peer phase locking of two X-band coaxial magnetrons using a 10 dB coupler and phase shifter, achieving an overall efficiency of 83.5% with an output of 309 MW and a pulse width of 3.5 μs [23]. Cheng *et al.* introduced a novel resonant cavity design for magnetrons, incorporating coupling bridges to enable mutual coupling among five magnetrons, ultimately achieving a relative phase jitter less than ±1° [24]. Li *et al.* utilized a 10 dB coupler to realize a hybrid injection scheme combining external and mutual injection, achieving a phase difference tuning range of −126° to 171° by adjusting the external signal frequency [25].

Peer-to-peer injection offers significant advantages: it effectively reduces the requirement for circulators and eliminates the requirement for external injection sources, significantly reducing insertion losses and system complexity. However, current methods either require modifications to the magnetron design or fail to eliminate the requirement for circulators and phase shifters, posing limitations for industrial applications.

The core components of MPC systems are H-plane tee combiners and T-junction power combiners. Since there is no additional isolation between the two input ports of H-plane tees, a certain level of coupling naturally occurs between them. This coupling characteristic enables using H-plane tees as a "coupling bridge" for peer-to-peer injection-locked magnetrons. A segment of waveguide is connected between the H-plane tee and one magnetron as a tuning part to adjust the system characteristics. The system requires neither any isolation components nor phase shifters. Therefore, a novel power combining system is proposed based on peer-to-peer locking.

Theoretical and numerical analyses of the peer-to-peer locking characteristics of the magnetrons are conducted. A qualitative explanation of how combining efficiency varies with different factors is presented. Then, a series of experiments is carried out to verify the proposed method. In the experiments, two 1-kW commercial continuous wave (CW) magnetrons were used for power combining. The experimental system achieved an optimal microwave efficiency of over 90%. Our analysis and experimental results demonstrate that high-efficiency power combining can be achieved only using the coupling characteristics of the combiner and commercially available magnetrons. The experimental system we built requires only an H-plane tee combiner and a section of waveguide to achieve efficient combining.

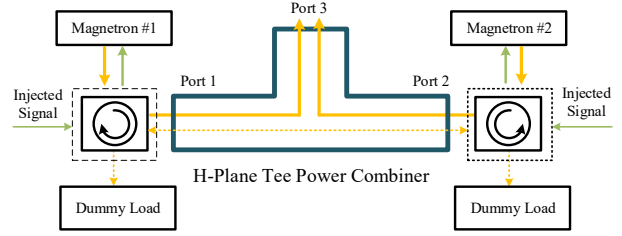

Fig. 1. Diagram of magnetron-based dual-way coherent power combining system.

This work leverages the coupling characteristics of the H-plane tee combiner to achieve high-efficiency power combining with peer-to-peer injection-locked magnetrons, offering advantages of compact size, low cost, and high efficiency. It provides a novel idea for developing MPC systems and broadens the applications of peer-to-peer injection-locked magnetrons. It also presents a novel solution to satisfy the current requirements of large-power microwave sources in industrial applications.

II. EFFICIENCY OF MAGNETRON POWER COMBINING SYSTEM

The efficiency of an MPC system is divided into combining efficiency $\eta_{com}$, microwave efficiency $\eta_{mw}$, and DC efficiency $\eta_{dc}$. The combining efficiency $\eta_{com}$ equals the system's output power divided by the total power input at the combiner ports. The microwave efficiency $\eta_{mw}$ is the output power divided by the total magnetron output power. The DC efficiency $\eta_{dc}$ is the ratio of the output power to the total DC power input. Their relationships are expressed as:

$$\begin{cases} \eta_{mw} = \eta_{com} \times \eta_{path} \\ \eta_{dc} = \eta_{mw} \times \eta_{MGT} \end{cases} \quad (1)$$

where $\eta_{MGT}$ represents the conversion efficiency of the magnetron, and $\eta_{path}$ is the path efficiency of the MPC system, which is associated with transmission losses. Typically, $\eta_{com}$ is a key index of the system performance. In this work, we primarily focus on the combining efficiency $\eta_{com}$ and microwave efficiency $\eta_{mw}$.

Fig. 1(a) presents a typical system diagram of a dual-way coherent power combining based on two magnetrons. The S-parameter matrix of the H-plane tee junction is:

$$[S_{H-Tee}] = \begin{bmatrix} |S_{11}|e^{j\alpha_1} & |S_{12}|e^{j\beta_1} & |S_{13}|e^{j\gamma_1} \\ |S_{21}|e^{j\beta_2} & |S_{22}|e^{j\alpha_2} & |S_{23}|e^{j\gamma_2} \\ |S_{31}|e^{j\sigma_1} & |S_{32}|e^{j\sigma_2} & 0 \end{bmatrix} \quad (2)$$

$V_1$ and $V_2$ represent the microwave amplitudes at the two input ports of the combiner, while $V_3$ is the microwave amplitude at the output port. $\theta_1$, $\theta_2$, and $\theta_3$ denote the phases of these three signals, respectively. Consequently, when the two signals pass through the H-plane tee, the signal conditions at each port are expressed as follows:

$$\begin{bmatrix} V_1^- \\ V_2^- \\ V_3^- \end{bmatrix} = [S_{H-Tee}] \begin{bmatrix} V_1^+ \\ V_2^+ \\ V_3^+ \end{bmatrix} = [S_{H-Tee}] \begin{bmatrix} |V_1^+|e^{j\theta_1} \\ |V_2^+|e^{j\theta_2} \\ 0 \end{bmatrix}$$

$$= \begin{bmatrix} |V_1^+||S_{11}|e^{j(\alpha_1+\theta_1)} + |V_2^+||S_{12}|e^{j(\beta_1+\theta_2)} \\ |V_2^+||S_{22}|e^{j(\alpha_2+\theta_2)} + |V_1^+||S_{21}|e^{j(\beta_2+\theta_1)} \\ |V_1^+||S_{31}|e^{j(\sigma_1+\theta_1)} + |V_2^+||S_{32}|e^{j(\sigma_2+\theta_1)} \end{bmatrix} \quad (3)$$

where $V_i^-$ and $V_i^+$ denote the input and output voltages at each port, respectively. Here, the H-plane tee is assumed to be ideal, so that $\sigma_1 = \sigma_2$, $|S_{31}| = |S_{32}| = -3$dB. Under these conditions, the combining efficiency of the MPC system shown in Fig. 1 is:

$$\eta_{com} = \frac{|V_1^+|^2 + |V_2^+|^2 + 2|V_1^+||V_2^+|\cos(\theta_2 - \theta_1)}{2(|V_1^+|^2 + |V_2^+|^2)} \times 100\%$$

$$= \left(\frac{1}{2} + \frac{\sqrt{K}}{1+K}\cos\theta\right) \times 100\% \quad (4)$$

where $K = |V_2^+|^2/|V_1^+|^2$ is the amplitude ratio of the two signals, $\theta = \theta_1 - \theta_2$ is the phase difference between the two signals.

To illustrate their relationships more intuitively, we first set $\theta$ to vary from 0 to 180° and $K$ to range from 0.1 to 10. The calculated $\eta_{com}$ are shown in Fig. 2(a). The efficiency of ideal coherent power combining reaches the highest value of 100% when $\theta = 0$ and $K = 1$. As the phase deviation $\theta$ increases and the power ratio $K$ grows, $\eta_{com}$ gradually decreases.

However, conventional injection-locked magnetrons usually require circulators to provide an injection path or phase shifters for phase control, as shown in Fig. 1. These components introduce extra insertion losses, making achieving the ideal system efficiency in practical systems challenging.

We denote the microwave losses of the two magnetron channels in the system as $ML$. The output amplitudes of the two magnetrons are $|V_{o1}|$ and $|V_{o2}|$, respectively. Thus, we will have $E_1 = |V_1|e^{j\theta_1} = ML \cdot |V_{o1}|e^{j\theta_1}$, $E_2 = |V_2|e^{j\theta_2} = ML \cdot |V_{o2}|e^{j\theta_2}$. Then, we take the $ML$ into consideration, and (4) becomes:

$$\eta_{MW} = ML^2 \left(\frac{1}{2} + \frac{\sqrt{K}}{1+K}\cos\theta\right) \times 100\% = ML^2 \eta_{com} \quad (5)$$

The insertion loss of each path also plays a significant role in the overall efficiency of a practical system. According to one estimation, the insertion loss of an isolation component, e.g., a circulator, is approximately 0.3 dB. Assuming all other components introduce an additional loss of 0.1 dB, we get $PL$ = 0.4 dB. Substituting this into (5), and setting $\theta$ varies from 0 to 180°, and $K$ from 0.1 to 10, the calculated $\eta_{MW}$ are presented in Fig. 2(b). The efficiency is significantly reduced, with the best microwave efficiency dropping from 100% to 91.2%. Thus, when the MPC system's path loss is too high, the system may still fall short of high-efficiency requirements even if an ideal combining efficiency is achieved. Reducing the system's path loss is crucial for achieving a high-efficiency power combining system.

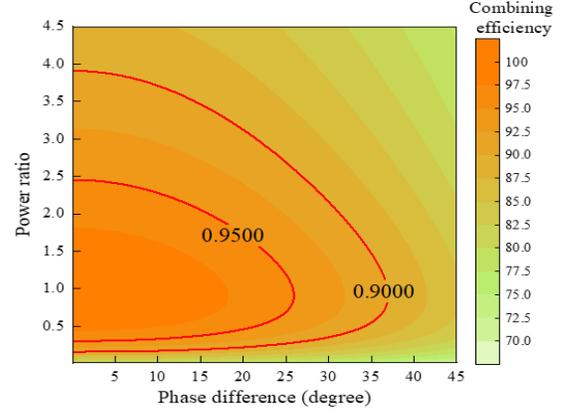

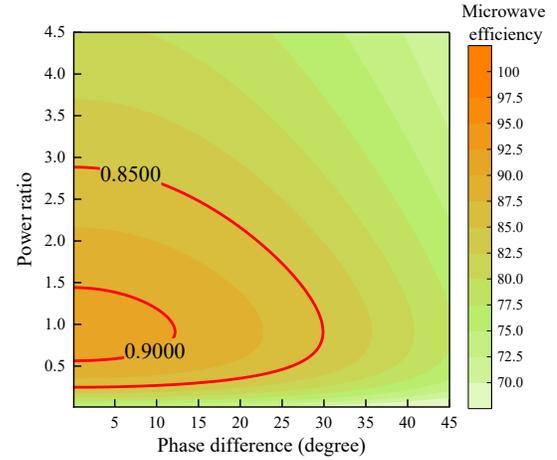

Fig. 2. Calculated (a) combining efficiency and (b) microwave efficiency of the MPC system in Fig. 1 with IL = 0.4 dB.

## III. THEORETICAL ANALYSIS

In previous works, H-plane tees have been widely used as power combiners. (3) indicates its characteristics among all ports, of which there is coupling between its two input ports. This characteristic allows the H-plane tee to serve as the power combiner and the "coupling bridge" between two magnetrons.

Peer-to-peer injection-locked magnetrons do not require any isolation components, making them an excellent solution for high-efficiency MPC systems. The coupling path between mutually injection-locked magnetrons typically requires

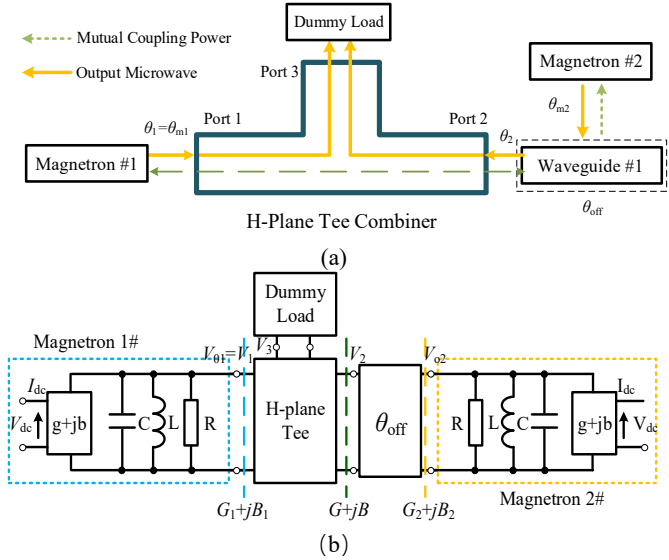

Fig. 3. (a) Diagram and (b) equivalent circuit schematic of proposed coherent power combining system.

careful design to minimize path loss and enhance system efficiency.

Therefore, we propose a novel design method for MPC systems, and the system diagram is shown in Fig. 3(a). The power flow and phase relationship are presented. In this system, a portion of each magnetron's output power is injected into the other through the H-plane tee, achieving phase locking. A phase compensation waveguide is connected between magnetron #2 and the H-plane tee to adjust the phase between the peer-to-peer locked magnetrons, with the corresponding phase shifts given by $\theta_{off}$. Therefore, the input phase $\theta_2$ at port 2 is defined as: $\theta_2=\theta_{m2}+\theta_{off}$, where $\theta_{m2}$ represents the output phase of the magnetron #2.

### A. Injection-Locking in the Proposed MPC System

According to the research by G. B. Collins and J. C. Slacter [26][27], a magnetron was modeled as a parallel RLC resonant network. The load and DC power supply influence its electron stream. The equivalent circuit schematic of the system is illustrated in Fig. 3(b). The R, L, and C in the diagram represent the equivalent resistance, inductance, and capacitance of the two magnetrons' resonant cavity, respectively. $V_{dci}$ and $I_{dci}$ refer to the two magnetrons' anode voltage and current supplied by the dc power source during the operation. $G_i+jB_i$ represents the magnetrons' equivalent load admittance.

We assume that the H-plane tee's right-side input port's admittance is $G+jB$. Since a section of transmission line is inserted between magnetron #2 and the H-plane tee, $G_2+jB_2$ is modified according to transmission line theory. Assuming the characteristic impedance of the transmission line is $Z_0$, we obtain:

$$G_2 + jB_2 = Z_{in} = Z_0 \frac{G + jB + jZ_0 \tan\theta_{off}}{Z_0 + j(G+jB)\tan\theta_{off}} \quad (6)$$

$G_2$=Re[$Z_{in}$], $B_2$=Im[$Z_{in}$]. Since the inserted waveguide segment shares identical characteristic impedance as the rest of the system, it only introduces a phase shift and does not affect the magnitude of the reflection coefficient.

In this case, the equilibrium oscillation requirement for Magnetron #i when operating independently is expressed as:

$$g_i = \frac{1}{R_i} + G_i, \quad b = \omega_{ci}C - \frac{1}{\omega_{ci}L} + B, \quad \omega_{0i}^2 = \frac{1}{LC} \quad (7)$$

where $\omega_{ci}$ and $\omega_{0i}$ represent the instantaneous output and resonant frequency of the $i$th magnetron resonant cavity, respectively.

According to J. C. Slater's work, we have:

$$g = \frac{1}{R}\left(\frac{V_{dc}}{V_{RF0}} - 1\right) \quad (8)$$

$$b = b_0 + g\tan\alpha \quad (9)$$

where $V_{RF0}$ is the amplitude of magnetrons' output power, $b_0$ and α are constants. Then, by solving (7), the free-running frequencies and output amplitude of the two magnetrons are:

$$\omega_i = \frac{b_i - B_i}{2C_i} + \omega_{0i}\sqrt{\frac{(b_i - B_i)^2}{4\omega_{0i}C_i^2} + 1} \quad (10)$$

$$V_{0i} = \frac{V_{dci}}{2R_iC_i\gamma_i}, \quad \gamma_i = \omega_{0i}\left(\frac{1}{Q_{0i}} + \frac{1}{2Q_{exti}}\right) \quad (11)$$

where $Q_{0i}=\omega_0CR$ and $Q_{exti}=\omega_0C/G$ represent the magnetrons' intrinsic and external quality factors, respectively.

When both magnetrons operate simultaneously, their oscillation states are no longer independent. Due to the coupling introduced by the H-plane tee, mutual coupling occurs between the output signals of the two magnetrons, as shown in Fig. 3(a). The amplitudes of the coupled signals at the two output ports are:

$$|V_{o1}^-| = |V_{o1}^+||S_{11}|e^{j(\alpha_1+\theta_{m1})} + |V_{o2}^+||S_{12}|e^{j(\beta_1+\theta_{m2}+\theta_{off})} \quad (12)$$

$$|V_{o2}^-| = |V_{o2}^+||S_{22}|e^{j(\alpha_2+\theta_{m2}+2\theta_{off})} + |V_{o1}^+||S_{21}|e^{j(\beta_2+\theta_{m1}+\theta_{off})} \quad (13)$$

From (3), it can be observed that the amplitudes of these two signals are determined by the inherent characteristics of the H-plane tee, the power of the two signals, and the phase difference between them. Here, the injection ratio is treated as a scalar quantity, accounting only for its magnitude. Accordingly, based on (12), (13) and the phase relationship of the system, the injection ratios of the two magnetrons are:

$$\rho_1 = \frac{|V_{o1}^-|}{|V_{o1}^+|} = K|S_{12}| + |S_{11}|\cos(\alpha_1 - \beta_1 - \Delta\theta - \theta_{off}) \quad (14)$$

$$\rho_2 = \frac{|V_{o2}^-|}{|V_{o2}^+|} = \frac{|S_{21}|}{K} + |S_{22}|\cos(\alpha_2 - \beta_2 + \Delta\theta + \theta_{off}) \quad (15)$$

where $\Delta\theta=\theta_{m2}-\theta_{m1}$.

In this case, the injection ratio is influenced by multiple factors. Specifically, $|S_{11}|$, $|S_{12}|$, $|S_{22}|$, $|S_{21}|$, $\alpha_1$, $\alpha_2, \beta_1$, and $\beta_2$ are intrinsic characteristics of the H-Plane Tee and are related to its design, $V_2^+$ and $V_1^+$ correspond to the output power of the magnetrons, $\theta_1$ and $\theta_2$ are associated with the final phase state of the system, and $\theta_{off}$ is related to our system design.

The effect on each magnetron is equivalently interpreted as a change in its load impedance when coupling power exists

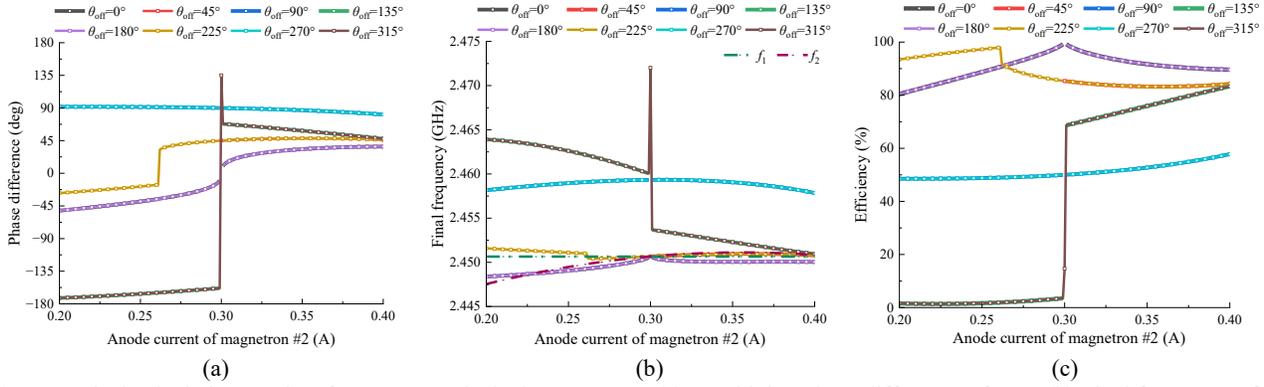

Fig. 4. Numerical calculation results of peer-to-peer locked magnetrons. (a) Combining phase difference $\Delta\theta_{com}$, (b) Final frequency of peer-to-peer locked magnetrons. (c) Calculated combining efficiency.

[27]. The resulting equivalent impedance is:

$$G' + jB' = \frac{|I_{oi}^+|e^{j\omega_i t} - |I_{oi}^-|e^{j\omega_{inj i} t}}{|V_{oi}^+|e^{j\omega_i t} + |V_{oi}^-|e^{j\omega_{inj i} t}} = \frac{|I_{oi}^+|}{|V_{oi}^+|} \left( \frac{1 + \frac{|I_{oi}^-|}{|I_{oi}^+|}e^{j(\omega_{inj i}-\omega_i)t}}{1 + \frac{|V_{oi}^-|}{|V_{oi}^+|}e^{j(\omega_{inj i}-\omega_i)t}} \right) \quad (16)$$

$$\approx G_i \left[1 - 2\rho_i e^{j\Delta\theta_i}\right] = G_i - 2G_i\rho_i \cos\Delta\theta_i - j2G_i \sin\Delta\theta_i$$

where $I_{oi}$ is the output current of each magnetron, $\omega_i$ represents the instantaneous output frequency under the influence of injected signals, and $\theta=(\omega_{inj}-\omega_c)t$ is the phase difference between the magnetron's output and injected signal. Then the equilibrium oscillation requirement will be changed as:

$$b_i = \omega_i C_i - \frac{1}{\omega_i L_i} - 2G_i \rho_i \sin\Delta\theta_i \quad (17)$$

Similarly, we assume $\omega_i \approx \omega_0$ solve (17), and subtract $\omega_{inj i}$ from both sides of it. Then we set $d\theta/dt = \omega_{inj i}-\omega_i$, the phase equations of the peer-to-peer locked magnetrons are obtained:

$$\frac{d\theta_{m1}}{dt} = \omega_1 - \omega_{inj 1} + \frac{\omega_{01}\rho_1}{Q_{ext1}} \sin\Delta\theta_1 \quad (18)$$

$$\frac{d\theta_{m2}}{dt} = \omega_2 - \omega_{inj 2} + \frac{\omega_{02}\rho_2}{Q_{ext2}} \sin\Delta\theta_2 \quad (19)$$

where $\omega_i$, $\omega_{0i}$, $\omega_{inj i}$, $\rho_i$, and $\Delta\theta_i$ ($i = 1,2$) represent the instantaneous output frequency, resonant frequency, injection frequency, injection ratio, and phase difference between the injection signal and the output signal of the $i$th magnetron, respectively. Here, we have the phase relationship: $\Delta\theta_1=\theta_{m2}-\theta_{m1}+\theta_{off}+\beta_1 = \Delta\theta+\beta_1+\theta_{off}$, $\Delta\theta_2=\beta_2+\theta_{off}+\theta_{m1}-\theta_{m2} = \beta_2+\theta_{off}-\Delta\theta_{com}$. Assuming $Q_{ext1}=Q_{ext2}$, when the system reaches a steady state, we have $d\theta/dt=0$, and $\omega_{inj1}=\omega_{inj2}=\omega_{final}$, $\omega_{01}=\omega_{02}=\omega_{0l}$. Under these conditions, (18) and (19) are rewritten as:

$$\omega_{final} - \omega_1 = \frac{\omega_0 \rho_1}{Q_{ext}} \sin(\Delta\theta + \beta_1 + \theta_{off}) \quad (20)$$

$$\omega_{final} - \omega_2 = \frac{\omega_0 \rho_2}{Q_{ext}} \sin(-\Delta\theta + \beta_2 + \theta_{off}) \quad (21)$$

By solving (20) and (21) simultaneously, we can derive the phase equation for peer-to-peer injection when the mutual coupling system reaches steady state, as well as the injection locking bandwidth:

$$\omega_1 - \omega_2 = \frac{\omega_0}{Q_{ext1}Q_{ext2}}[Q_{ext1}\rho_2 \sin(\beta_2 + \theta_{off} - \Delta\theta) \quad (22)$$

$$-Q_{ext2}\rho_1 \sin(\Delta\theta + \beta_1 + \theta_{off})]$$

(22) indicates that the phase difference $\Delta\theta_{com} = \theta_2-\theta_1 = \Delta\theta+\theta_{off}$ between the two input ports of the H-plane tee. The locking bandwidth is adjusted by multiple factors. Some of these are related to the output characteristics of the magnetrons, such as their output frequency, output power, and the injection ratios between them.

To provide a more intuitive representation of the solution, (22) has been reformulated using trigonometric identities as follows (with details of the derivation demonstrated in Appendix A):

$$M + R\sin(\Delta\theta + \phi) = 0 \quad (23)$$

where

$$M = Q_{ext1}Q_{ext2}\frac{\omega_2 - \omega_1}{\omega_0}, \; \phi = \arctan\left(\frac{A}{B}\right), \; R = \sqrt{A^2 + B^2} \quad (24)$$

$$A = Q_{ext1}\rho_2 \sin(\beta_2 + \theta_{off}) - Q_{ext2}\rho_1 \sin(\beta_1 + \theta_{off}) \quad (25)$$

$$B = -Q_{ext1}\rho_2 \cos(\beta_2 + \theta_{off}) - Q_{ext2}\rho_1 \cos(\beta_1 + \theta_{off}) \quad (26)$$

Provided that $|M/R| \leq 1$, (23) is rewritten as:

$$\Delta\theta = -\phi - \arcsin\left(\frac{M}{R}\right), \; or \; \Delta\theta = -\phi + \pi + \arcsin\left(\frac{M}{R}\right) \quad (27)$$

$|M/R|\leq 1$ represents the locking condition. (27) reveals two possible solutions, which depend on $M$, $R$, and $\phi$ — i.e., on the coupling between the two magnetrons, their injection ratios, the combiner's phase characteristic, and the waveguide length $\theta_{off}$. The solution is stable when it satisfies $\cos(\phi+\Delta\theta)<0$. It is important to recognize that the phase difference $\Delta\theta$ appears explicitly in the trigonometric functions and the coupling parameters (e.g., $\rho_1$, $\rho_2$), and both $\rho_1$ and $\rho_2$ are also functions of $\Delta\theta$, forming an implicit nested relationship. Consequently, (27) is not an explicit analytic expression for $\Delta\theta$ but rather an alternative formulation of (23). Additionally, $M$, $R$, and $\phi$ are influenced by $\Delta\theta$ and these variables interact until a steady state is reached.

Furthermore, the existence of multiple solutions implies that when the injection conditions vary, the system may abruptly transition to another solution that satisfies the locking condition, rather than evolving continuously. This work concentrates on how $\Delta\theta$ is tuned by adjusting system design and magnetron output characteristics. Hence, a more in-depth discussion of the injection characteristics is not pursued here.

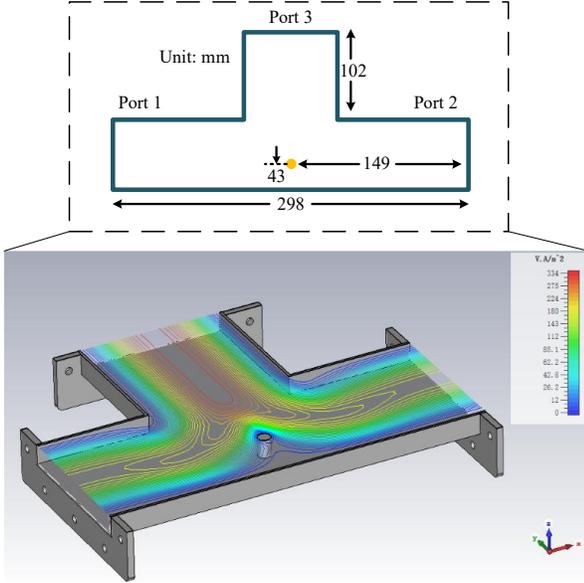

Fig. 5.  Physical dimension and power flow of fabricated H-plane tee combiner.

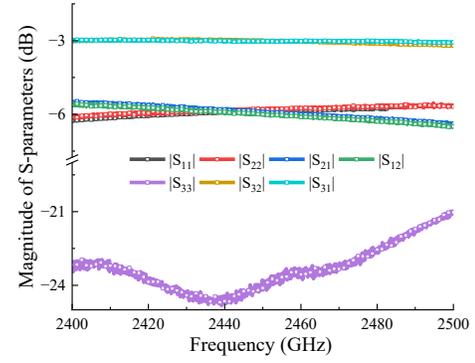

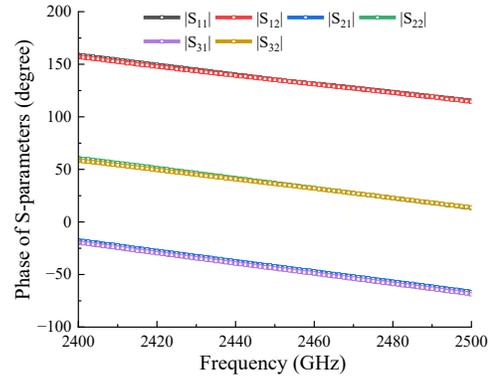

Fig. 6.  Measured (a) magnitude and (b) phase of the H-plane tee combiner's S-parameters.

From the foregoing, we conclude that by appropriately tuning relevant system parameters, it is possible to achieve high-efficiency coherent power combining within the system. To further illustrate how the system output characteristics respond under different conditions, a series of numerical analyses is conducted to qualitatively evaluate the influence of various parameters.

### B. Qualitative Numerical Analysis

We conducted a series of calculations to illustrate the influence of $\theta_{\text{off}}$ and magnetron output on the final combined phase. According to the analysis in the previous section, an explicit analytical solution for $\Delta\theta$ is unattainable. Therefore, we employ a numerical iterative algorithm to solve the equation, yielding high-precision results and allowing for a more intuitive analysis and presentation of the locking phase and frequency characteristics of the proposed system.

As shown in (10) and (11), both output frequency and power of a magnetron are closely associated with its power supply and load impedance, which is commonly known as frequency pushing and pulling effects. To facilitate the calculation of the magnetrons' output characteristics in (14), we temporarily neglect the frequency pulling effects and instead account for frequency pushing by adopting the polynomial fits, derived in our previous work[28], that relate each magnetron's output frequency and power to its anode current：

$$f_c(I_{dc}) = a_0 + a_1 I_{dc} + a_2 I_{dc}^2 \qquad (28)$$

where $a_0$=2.4324, $a_1$=0.1053, $a_2$= −0.1484, and:

$$P_{RF0}(I_{dc}) = b_0 + b_1 I_{dc} + b_2 I_{dc}^2 \qquad (29)$$

where $b_0$= −41.9230, $b_1$=2748.9220, $b_2$= −294.9274. $I_{dc}$ refers to the magnetron anode current.

In this case, (22) is solved by a numerical iterative algorithm. We have the following settings: $Q_{\text{ext1}} = Q_{\text{ext2}} = 50$, $|S_{11}| = -6$ dB, $|S_{12}|=-6$ dB, $|S_{22}| = -6$ dB, $|S_{21}| = -6$ dB, $\alpha_1 = 135°$, $\alpha_2 = 135°$, $\beta_1 = -45°$, $\beta_2 = -45°$. Here, we set the current of magnetron #1 to 0.3 A, while varying the current of magnetron #2 from 0.2 A to 0.4 A to observe the final combined phase and frequency under different conditions. $\theta_{\text{off}}$ is set from 0° to 315° in steps of 45°. Additionally, in this calculation, the effect of load pulling on the magnetron frequency is neglected due to the difficulty of fitting and quantifying this characteristic. It is assumed that the two magnetrons follow the same output behavior. We substitute the calculated $\Delta\theta$ into $\Delta\theta_{\text{com}}=\Delta\theta+\theta_{\text{off}}$ and then input the corresponding power and phase values into (4) in order to obtain the phase difference at the H-plane tee's input port and the corresponding combining efficiency.

The calculated results are shown in Fig. 6. The points where a steady-state solution cannot be obtained are not shown in the figure. Fig. 4(a) and (b) illustrate how $\Delta\theta_{\text{com}}$ and $f_{\text{final}}$ vary with the current of magnetron #2 for different $\theta_{\text{off}}$. It shows that varying $\theta_{\text{off}}$ effectively adjusts the steady-state output phase of the mutually injection-locked magnetrons, thereby controlling the phase difference of the input signals to the H-Plane tee combiner and achieving high-efficiency power combining. Additionally, different $\theta_{\text{off}}$ values lead the peer-to-peer magnetrons to operate at different steady-state frequencies.

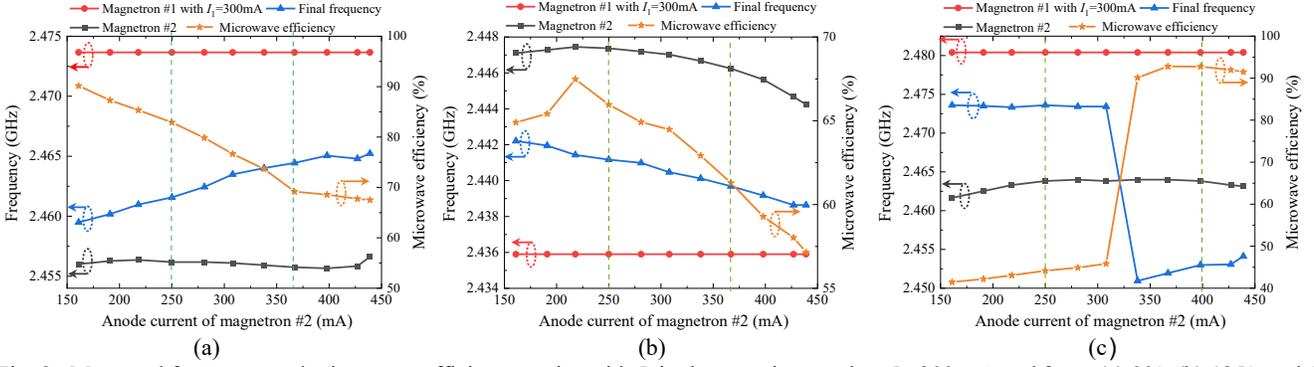

Fig. 8. Measured frequency and microwave efficiency varies with $I_2$ in the experiment when $I_1$=300 mA and $\theta_{off}$ = (a) 90°, (b) 135°, and (c) 225°.

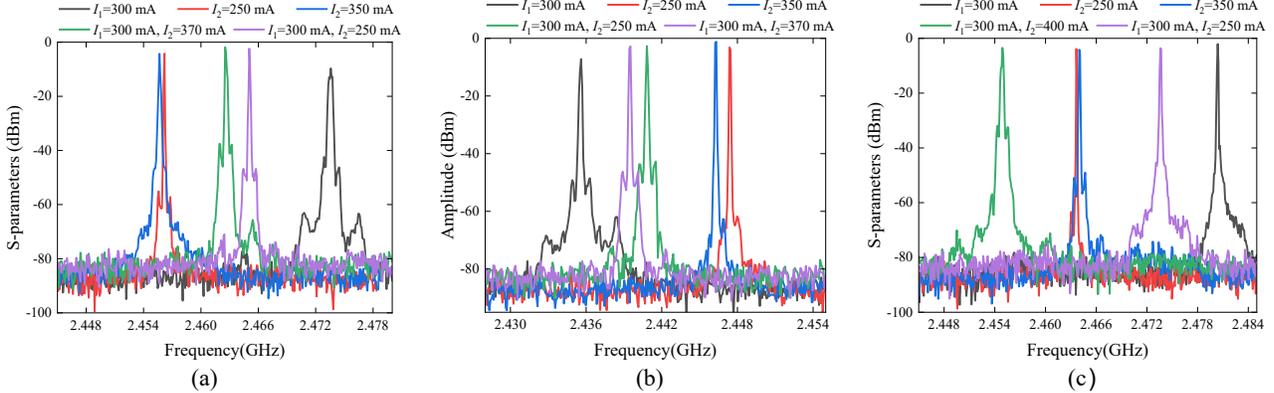

Fig. 9. Spectra of single magnetron output and peer-to-peer locked output under selected conditions when $\theta_{off}$ = (a) 90°, (b) 135°, and (c) 225°. (RBW: 30 kHz, VBW: 30 kHz)

Fig. 4(c) shows the corresponding combining efficiency, and the efficiency varies with the phase difference in Fig. 6(a). This result indicates that by adjusting $\theta_{off}$ and the output characteristics of the magnetrons, the system can be driven to operate in a state with high combining efficiency.

At certain values of $\theta_{off}$ as the anode current of magnetron #2 varies, the peer-to-peer injection-locked magnetrons exhibit two distinct locking modes, corresponding to different phase differences and final frequencies. This phenomenon arises from the variations in power and frequency differences between the two magnetrons, which in turn change the locking conditions, eventually causing the magnetrons cannot be locked in the original state and forcing a jump to an alternative solution of the equation. We refer to these distinct states as different locking modes.

Based on these results, the high-efficiency power combining in peer-to-peer injection-locked magnetron systems is achieved by appropriately choosing $\theta_{off}$ and tuning the operating conditions of the two magnetrons

IV. EXPERIMENTAL VERIFICATION AND RESULTS

A. *H-Plane Tee Combiner*

An H-plane tee combiner was designed and fabricated using WR430 waveguides with a metal post. Fig. 5 shows its internal power flow diagram and specific dimensions. The measured S-parameters and phase responses of the combiner's ports are presented in Fig. 6(a) and 6(b), respectively.

Here, we primarily focus on its performance around

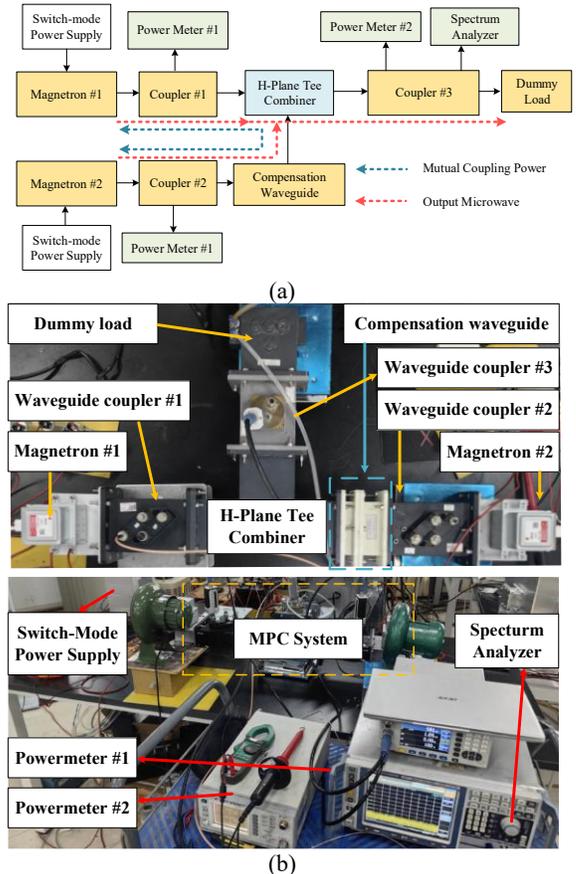

Fig. 7. Setup of the experiment system. (a) Diagram of the experiment system. (b) Photograph of the experiment system.

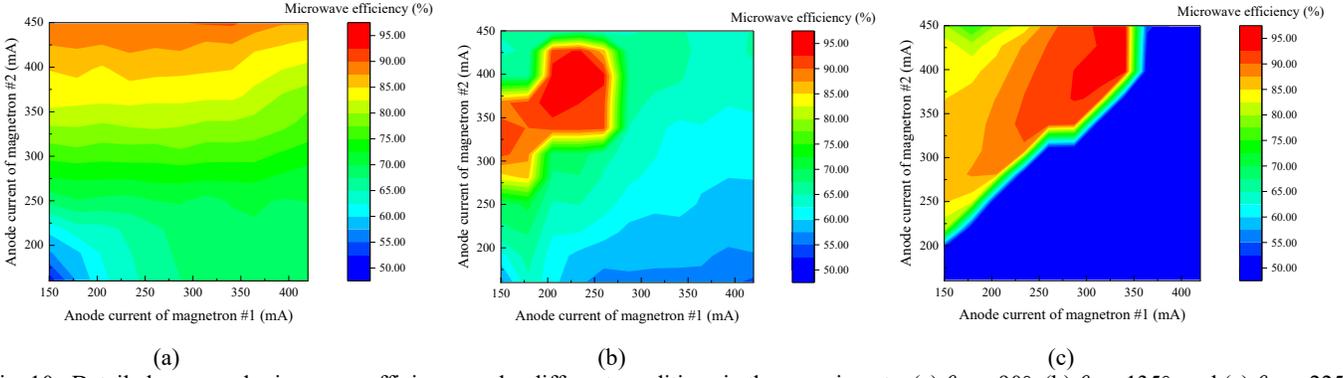

Fig. 10. Detailed measured microwave efficiency under different conditions in the experiments. (a) $\theta_{off}$ = 90°, (b) $\theta_{off}$ =135°, and (c) $\theta_{off}$ = 225.

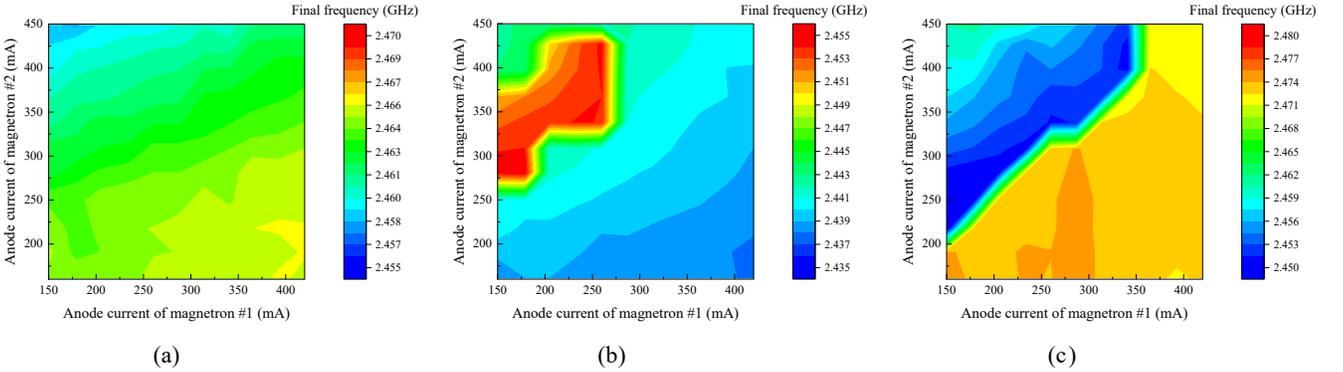

Fig. 11. Detailed measured final frequency under different conditions in the experiments. (a) $\theta_{off}$= 90°, (b) $\theta_{off}$ =135°, and (c) $\theta_{off}$ = 225.

2.45 GHz. At 2.45 GHz, $|S_{21}|$ and $|S_{12}|$ are –5.97 dB and –6.03 dB, respectively; $|S_{11}|$ and $|S_{22}|$ are –5.87 dB and –5.86 dB, respectively; and $|S_{33}|$ is –24 dB. Moreover, for the phase characteristic of the H-plane tee, $\alpha_1$ = 135.4°, $\alpha_2$ = 135.1°, $\beta_1$ = 44.8°, $\beta_2$ = 43.1°, $\sigma_1$ = 36.3°, and $\sigma_2$ = 36.9°. These results meet the performance requirements for a standard H-plane tee combiner.

### B. Experiment Setup

An experimental system is established to verify the proposed method. The block diagram of the magnetron experimental system used in this paper is shown in Fig. 7(a). Fig. 7(b) shows two photographs of the system. Two 1-kW S-band CW magnetrons (LG, 2M244) are powered by two switch-mode CW power supplies (WELAMP 2000F, Magmeet), respectively. These power supplies can adjust the magnetrons' anode currents, thereby tuning the output power and frequency. A capacitor filter module has been applied to improve the switch-mode power supply. The filament current supply is cut off automatically after the magnetron works normally.

In the experiment, several $\theta_{off}$ values are selected to illustrate the influence of $\theta_{off}$ on the magnetrons' final steady state. Three values of $\theta_{off}$ (90°, 135°, and 225°), which have distinct characteristics as shown in Fig. 4, are selected considering the installation constraints and system size. It should be noted that measurements were also attempted at $\theta_{off}$ = 0°. However, the combining efficiency dropped to nearly 0% under this condition. To avoid potential damage from the resulting strong standing waves to both the magnetrons and the measurement equipment, extended testing and detailed data collection at $\theta_{off}$ = 0° were not conducted. Consequently, these results are omitted from this paper.

Magnetron #1 is directly connected to the H-plane tee designed in Section III, while magnetron #2 is connected to the other input port through a compensation waveguide. The output spectrum of the magnetron is measured by a spectrum analyzer (FSP, R&S), and the port power of the H-plane tee is measured by two power meters (AV2433 and 2438CB, Ceyear). A water-cooled dummy load is used to absorb the microwave power output from the magnetrons.

In the experiment, the anode current varies from approximately 150 mA to 450 mA to ensure stable operation of the magnetrons. The anode currents of the two magnetrons are represented as $I_1$ and $I_2$, respectively. In addition, due to the inherent frequency jitter in peer-to-peer injection-locked magnetrons, it is difficult to accurately measure the phase difference between them. Therefore, only the microwave efficiency of the system was measured in the experiment to reflect the variation in phase difference and the efficiency of power combining.

### A. Results and Discussion

Firstly, based on the setup used in the numerical analysis, we conducted a series of measurements on the experimental system. Specifically, the output frequency of magnetron #1 was measured under the condition of $I_1$ = 300mA when operating independently. Subsequently, $I_2$ varied from 150 mA to 450 mA, also under independent operation. Then, the corresponding injection-locked final frequency and the microwave efficiency of coherent power combining were measured to evaluate the impact of the magnetron's anode

TABLE I
MEASURED MICROWAVE EFFICIENCY UNDER DIFFERENT CONDITIONS

| Compensation | Magnetron#1 | | Magnetron#2 | | Total Output | |
|---|---|---|---|---|---|---|
| $\theta_{\text{off}}$ | Output Power | Anode Current | Output power | Anode Current | Total Power | $\eta_{\text{MW}}$ |
| 90° | 940 W | 363 mA | 1070 W | 449 mA | 1790 W | 89.1% |
|  | 854 W | 341 mA | 1050 W | 448 mA | 1720 W | 90.3% |
|  | 780 W | 312 mA | 1050 W | 448 mA | 1650 W | 90.2% |
|  | 638 W | 290 mA | 1040 W | 448 mA | 1510 W | 90.0% |
|  | 498 W | 207 mA | 1030 W | 448 mA | 1370 W | 89.7% |
| 135° | 570 W | 260 mA | 845 W | 398 mA | 1320 W | 93.3% |
|  | 500 W | 232 mA | 846 W | 398 mA | 1260 W | 93.6% |
|  | 458 W | 209 mA | 720 W | 337 mA | 1080 W | 91.6% |
|  | 320 W | 152 mA | 730 W | 338 mA | 950 W | 90.4% |
|  | 335 W | 153 mA | 632 W | 308 mA | 878 W | 90.8% |
| 225° | 781 W | 339 mA | 1010 W | 448 mA | 1670 W | 93.2% |
|  | 782 W | 339 mA | 938 W | 392 mA | 1610 W | 93.6% |
|  | 718 W | 312 mA | 874 W | 363 mA | 1480 W | 93.0% |
|  | 533 W | 251 mA | 756 W | 340 mA | 1170 W | 90.8% |
|  | 535 W | 251 mA | 692 W | 311 mA | 1100 W | 89.6% |

TABLE II
COMPARISON OF PHASE STABILITY OF OUR PROPOSED METHOD WITH OTHER POWER COMBINING METHODS

| | Work | Best Combining Efficiency | Best Microwave Efficiency | Output Power at Best Efficiency | Isolation | Injection-locking | Injection Power | Phase Noise |
|---|---|---|---|---|---|---|---|---|
| | [15] | 97.7% | NM | 1.42 kW | Four circulators | External | 33 W | NM |
| | [18] | 94.5% | 94.5% | 1.46 kW | One magic-tee | External | 1×26 W | NM |
| | [19] | 93% | NM | 2.45 kW | Two circulators and two magic-tees | External | 4×13 W | NM |
| | [20] | 92% | NM | NM | One circulator | Master-Slave | NA | NM |
| | [29] | 94.7% | 86.7% | 34.0 kW | Two circulators | Asymmetric | 2×100 W | −65.0 dBc@500 kHz |
| | [30] | 95.7% | NM | 1.57 kW | One circulator | External and Master-Slave | 1×15 W | −61.0 dBc@10 Hz, −80.9 dBc@ 100 Hz, −91.6 dBc@ 1 kHz. |
| | [31] | 94.6% | NM | 1.89 kW | Two circulators | External Injection Locking | 2×72 W | NM |
| This work | $\theta_{\text{off}} = 90°$ | 90.3% | 90.3% | 1.72 kW | None | Peer-to-peer (no injection) | NA | -62.52dBc@500kHz |
| | $\theta_{\text{off}} = 135°$ | 93.6% | 93.6% | 1.26 kW | None | Peer-to-peer (no injection) | NA | -66.74dBc@500kHz |
| | $\theta_{\text{off}} = 225°$ | 93.6% | 93.6% | 1.61 kW | None | Peer-to-peer (no injection) | NA | -60.21dBc @500kHz |

*NM: not mentioned, NA: not applicable.

current on peer-to-peer injection locking performance under different values of $\theta_{\text{off}}$.

Fig. 8 shows that the measured frequency and microwave efficiency vary with $I_2$ in the experiment when $I_1$=300 mA. The output frequency of the magnetrons when operating individually varies with $\theta_{\text{off}}$, which is attributed to their frequency pulling characteristics as shown in (10) and (11). A comparison of Fig. 8(a)–(c) reveals that under different values of $\theta_{\text{off}}$, the variation trends of the peer-to-peer injection-locked magnetrons' output frequency and the system's microwave efficiency with respect to the anode current of magnetron #2 differ noticeably, which is consistent with the observations from the numerical calculation results.

It also shows that the final oscillation frequency and microwave efficiency vary smoothly as $I_2$ increases for $\theta_{\text{off}} = 90°$ and 135°. In contrast, for $\theta_{\text{off}} = 225°$, both the frequency and efficiency undergo an abrupt transition around $I_2$=360 mA, This behavior arises because changing $\theta_{\text{off}}$ influences the system's steady-state condition. As the power imbalance between the two magnetrons grows, mutual injection can no longer sustain the original frequency and phase, forcing a jump into a new locking state. Moreover, for $\theta_{\text{off}} = 90°$ and 225°, higher steady-state frequencies correspond to lower microwave efficiencies—implying larger phase differences—whereas at $\theta_{\text{off}} = 135°$, efficiency increases with frequency. These observations confirm that, under peer-to-peer injection locking, microwave efficiency and the final frequency shift are simultaneously consistent with both our theoretical derivations and numerical calculations. Variations of $\theta_{\text{off}}$ or the magnetrons' output characteristics will shift the phase difference and final frequency.

Subsequently, we measured the spectra under different conditions, as shown in Fig. 9(a)–(b), which illustrate the frequency responses at the dashed positions in Fig. 8(a)–(c) for both individual magnetron operation and peer-to-peer injection-locked. Once both magnetrons are operating, injection locking occurs, and their spectra are pulled toward a new common frequency. Furthermore, as magnetron #2 is adjusted, the final locked frequency also shifts accordingly. In the experiments, the locking state remained stable and persistent unless the magnetron operating conditions were changed by tuning the anode current, causing the system to transition into another locking state.

Fig. 10 and Fig. 11 provide a more detailed experimental analysis of the microwave efficiency and frequency characteristics of the power combining system. These results offer a clearer demonstration of how $\theta_{off}$ and the anode currents of the two magnetrons influence the system's performance. When $\theta_{off}$ varies, the optimal combining efficiency and the final locking frequency of the system change accordingly.

When $\theta_{off} = 90°$, the system exhibits a smooth transition as the anode currents of the two magnetrons are adjusted, with $I_2$ having a more pronounced influence on the system behavior. The highest microwave efficiency in this case is 90.3%, occurring at $I_1 = 341$ mA and $I_2 = 449$ mA, with a final frequency of 2.4634 GHz. However, when $\theta_{off} = 135°$ and 225°, the system behaves differently: adjusting $I_1$ and $I_2$ clearly leads to two distinct injection-locked states, and one of the locking states corresponds to significantly higher microwave efficiency, as shown in Fig. 4(c). One state yields higher microwave efficiency (i.e., lower phase difference), while the other corresponds to lower microwave efficiency. For $\theta_{off} = 135°$, the optimal efficiency reaches 93.7% at $I_1 = 205$ mA and $I_2 = 400$ mA, with a final frequency of 2.45143 GHz. When $\theta_{off} = 225°$, the highest efficiency is 93.6%, occurring at $I_1 = 341$ mA and $I_2 = 398$ mA, with the final frequency of 2.45109 GHz. At $\theta_{off} = 225°$, the distinction between the two locking states becomes more pronounced, and the region of high-efficiency operation is broader compared to the case at 135°. These results indicate that $\theta_{off}$ significantly influences the system behavior.

Fig. 8 to 11 also demonstrate that tuning the output characteristics of the magnetrons can effectively control the system's final phase difference and frequency. While the experimental results show some quantitative discrepancies compared to the numerical calculations, this is primarily due to simplifications in the numerical model, which neglects load pulling characteristics and differences in factors such as external quality factors between the two magnetrons. Nevertheless, the overall trends observed in the simulations are qualitatively consistent with those seen in the experiments, supporting the validity of the theoretical insights.

Table I presents a part of the measured microwave efficiency under different $\theta_{off}$. The system can consistently achieve over 90% microwave efficiency in various scenarios. However, the output power achievable at high efficiency is different. The system achieves its optimal microwave efficiency of 90.3% with an output power of 1720 W when $\theta_{off}$ is 90°. It maintains a microwave efficiency of over 90% within the output power range of 1510–1720 W. At $\theta_{off} = 135°$, the output power corresponding to high microwave efficiency is lower, with over 90% efficiency achieved between 878 W and 1320 W, and the highest efficiency of 93.6% observed at 1260 W. For $\theta_{off}$ is 225°, the system maintains a microwave efficiency of over 90% within the output power range of 1170–1670 W, achieving the highest microwave efficiency of 93.6% at 1260 W. These results demonstrate the effectiveness of the proposed method. By designing the experimental system appropriately, high-efficiency power combining is achieved using peer-to-peer injection-locked magnetrons without any isolation components, injection signals, or phase shifters.

Table II compares the system in our work with previous work. Unlike other approaches, our method requires no isolation components or injection signal source, significantly reducing system cost and size. Due to the absence of an external highly stable injection signal and precise phase adjustment via phase shifters, the proposed method exhibits slightly lower combining efficiency than systems employing external injection. In addition, its phase noise is relatively worse compared to systems with external injection, and the output contains more spurious components. However, it offers significant advantages over other systems without external injection sources. In addition, compared with external injection-locking systems, our method does not require isolators or injection signal sources, greatly reducing the system's cost and size. Furthermore, by minimizing system losses, the proposed method achieves higher microwave efficiency than the method with higher combining efficiency [29]. The advantages of low loss, low cost, and compact size make this method suitable for industrial microwave applications, providing a new perspective for meeting the high-power microwave demands in industrial settings.

## V. CONCLUSION

A novel method for achieving high-efficiency coherent power combining of magnetrons is proposed. Without requiring redesign of the magnetron structure, the method enables efficient combining using only commercial magnetrons and an H-plane tee combiner. The coupling characteristics of the H-plane tee combiner are utilized to realize peer-to-peer injection locking between two magnetrons. The injection locking characteristic and the associated phase equations are derived and qualitatively analyzed. Finally, an MPC system based on this method is designed, constructed, and experimentally validated.

In the experiments, by using a section of waveguide to tuning the system's phase characteristic, the proposed system achieved maximum microwave efficiencies of 90.2%, 93.6%, and 93.6% at waveguide electrical lengths $\theta_{off}= 90°$, 135°, and 225°, with output powers of 1650 W, 1260 W, and 1610 W, respectively. By adjusting the anode currents of the two magnetrons, the system maintained over 90% microwave efficiency within the output power ranges of 1510–1720 W, 878–1320 W, and 1100–1670 W, respectively. These experimental results demonstrate the effectiveness of the proposed method. Moreover, the observed trends in microwave efficiency and output frequency under varying $\theta_{off}$ and magnetron anode currents are consistent with those predicted by theoretical and numerical analysis.

This method is expected to be applied in developing novel power combining systems for applications such as microwave chemistry, microwave heating, and wireless microwave power transmission, where stringent signal quality is not required. It enables the realization of compact, low-cost, and high-power microwave sources. Furthermore, the proposed method may be extended to N-way power-combining systems (configuration given in Appendix B). This study offers a new approach for efficient power combining with a simple system architecture, small size, and low cost. It is anticipated that further advancement will be made in applying peer-to-peer injection-locked magnetrons in power combining systems.

## APPENDIX A

For the derivation of (23), (22) is firstly decomposed into two parts: a linear term $M$ and a nonlinear term $N$, as given by:

$$M = Q_{ext1} Q_{ext2} \frac{\omega_2 - \omega_1}{\omega_0} \tag{30}$$

$$N = Q_{ext1}\rho_2 \sin(\beta_2 + \theta_{off} - \Delta\theta) \\ - Q_{ext2}\rho_1 \sin(\Delta\theta + \beta_1 + \theta_{off}) \tag{31}$$

By applying trigonometric identities, the nonlinear term $N$ will be rewritten as:

$$N = Q_{ext1}\rho_2 \left[ \sin(\beta_2 + \theta_{off})\cos(\Delta\theta) - \cos(\beta_2 + \theta_{off})\sin(\Delta\theta) \right] \\ - Q_{ext2}\rho_1 \left[ \sin(\theta_{off} + \beta_1)\cos(\Delta\theta) + \cos(\theta_{off} + \beta_1)\sin(\Delta\theta) \right] \tag{32}$$

Therefore, by combining the polynomial terms, we obtain:

$$N = \sin(\Delta\theta)\left[-Q_{ext2}\rho_1 \cos(\theta_{off} + \beta_1) - Q_{ext1}\rho_2 \cos(\beta_2 + \theta_{off})\right] \\ + \cos(\Delta\theta)\left[Q_{ext1}\rho_2 \sin(\beta_2 + \theta_{off}) - Q_{ext2}\rho_1 \sin(\theta_{off} + \beta_1)\right] \tag{33}$$

Here, we have:

$$A = Q_{ext1}\rho_2 \sin(\beta_2 + \theta_{off}) - Q_{ext2}\rho_1 \sin(\beta_1 + \theta_{off}) \tag{34}$$

$$B = -Q_{ext1}\rho_2 \cos(\beta_2 + \theta_{off}) - Q_{ext2}\rho_1 \cos(\beta_1 + \theta_{off}) \tag{35}$$

$$\phi = \arctan\left(\frac{A}{B}\right), \quad R = \sqrt{A^2 + B^2} \tag{36}$$

Then, $N$ will be rewritten as:

$$N = A\cos(\Delta\theta) + B\sin(\Delta\theta) \tag{37}$$

Therefore, by substituting (30) and (37) into (22), and combined with (34) — (36), the (23) will be obtained:

$$M + R\sin(\Delta\theta + \phi) = 0 \tag{38}$$

## APPENDIX B

For an N-way ($N = 2^m$, $m = 1,2,3,\ldots$) coherent power combining system based on this method, we propose a possible system configuration. Here, we do not specify the exact type of power combiner used (such as the combiner formed by combining H-plane and E-plane tees as in [14] or the independent five-port combiner in [15]); instead, we simply assume that the combiner is a lossless ($N$+1)-port power combining device. If the output port is ideally matched, such a device inherently exhibits certain coupling between its input ports.

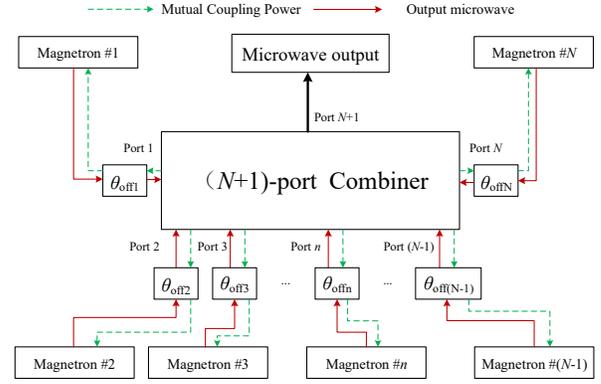

Fig. 12.  Diagram of a N-way power combining system.

Based on the system we proposed earlier, each output port of the combiner will be connected to a section of phase compensation waveguide to balance the phase differences among the magnetrons. Fig. 12 illustrates the block diagram of this $N$-way system. The ports 1 through $N$ serve as the input ports, while port $N$+1 acts as the output port. For this network, its S-parameter matrix is expressed as:

$$[S_{N-combiner}] = \begin{bmatrix} |S_{11}|e^{j\theta_{11}} & |S_{12}|e^{j\theta_{12}} & \ldots & |S_{1(N+1)}|e^{j\theta_{1(N+1)}} \\ |S_{21}|e^{j\theta_{21}} & |S_{22}|e^{j\theta_{22}} & \ldots & \ldots \\ \ldots & \ldots & \ldots & \ldots \\ |S_{(N+1)1}|e^{j\theta_{(N+1)1}} & \ldots & \ldots & |S_{(N+1)(N+1)}|e^{j\theta_{(N+1)(N+1)}} \end{bmatrix} \tag{39}$$

Assume the compensation waveguide connected to port $n$ (where $n$ = 1, 2, …, $N$) have an electrical length of $\theta_{off, n}$, the output phase of magnetron #$n$ be $\theta_n$ and the output amplitude be $V_n$. Then, based on the derivation presented in Sec. II, under these conditions, the signal at port $n$ and the corresponding injection ratio can be expressed as:

$$V_n^- = \sum_{i=1}^{i=N} |V_i^+||S_{ni}|e^{j(\theta_{ni} + \theta_i + \theta_{off,i} + \theta_{off,n})} \tag{40}$$

$$\rho_1 = \frac{|V_n^-|}{|V_n^+|} \tag{41}$$

Then, the injection-locking equations for this system will be extended into the following set of equations:

$$\begin{cases} \omega_{final} - \omega_1 = \frac{\omega_0 \rho_1}{Q_{ext}} \sin(\theta_{inj,1}) \\ \omega_{final} - \omega_2 = \frac{\omega_0 \rho_2}{Q_{ext}} \sin(\theta_{inj,2}) \\ \ldots \\ \omega_{final} - \omega_n = \frac{\omega_0 \rho_n}{Q_{ext}} \sin(\theta_{inj,n}) \\ \ldots \\ \omega_{final} - \omega_N = \frac{\omega_0 \rho_N}{Q_{ext}} \sin(\theta_{inj,N}) \end{cases} \tag{42}$$

where $\theta_{inj,n}=\theta_{inj}-\theta_n$, $\theta_{inj}$ is the phase of the injection signal at magnetron #$n$'s port, i.e., phase of $V_n^-$.

According to (40), (41) and (42), the final steady state of the system depends on the output characteristics of each magnetron as well as the lengths of the compensation waveguides. By solving (42), the characteristics of the N-way system can be obtained. When $N = 2$, (40), (41) and (42) degenerate into the form given in Sec. II. A more detailed analysis and calculation for the N-way system is not included in this paper and will be investigated in our future work.

## REFERENCES

[1] S. Ud-Din Khan et al., "Cost-Effective Microwave Assisted ECR Heating Using Combination of Quasi-Locked Low-Power Magnetrons on GLAST-III," in *IEEE Plasma Sci.*, vol. 52, no. 6, pp. 2021-2028, June 2024, doi: 10.1109/TPS.2024.3443132.

[2] P. Anilkumar, A. Kumar, D. Pamu and T. Tiwari, "Design and Thermal Study of 5 MW S-Band Tunable Pulsed Magnetron for Linear Accelerator System," in *IEEE Plasma Sci.*, vol. 51, no. 5, pp. 1223-1231, May 2023, doi: 10.1109/TPS.2023.3264813.

[3] B. Yang, X. Chen, J. Chu, T. Mitani, and N. Shinohara, "A 5.8-GHz Phased Array System Using Power-Variable Phase-Controlled Magnetrons for Wireless Power Transfer," in *IEEE Trans. Microw. Theory Techn.*, vol. 68, no. 11, pp. 4951-4959, Nov. 2020, doi: 10.1109/TMTT.2020.3007187.

[4] S. Wang, Y. Shen, C. Liao, J. Jing, and C. Liu, "A Novel Injection-Locked S-Band Oven Magnetron System Without Waveguide Isolators," in *IEEE Trans. Electron Devices*, vol. 70, no. 4, pp. 1886-1893, Apr. 2023, doi: 10.1109/TED.2023.3243041.

[5] H. Zhang, L. Chen, X. Chen, Y. Liao and S. Wang, "Feedback Control for Stable Output Power of Magnetrons Based on Signal Reconstruction," in *IEEE Trans. Microw. Theory Techn.*, vol. 72, no. 8, pp. 4508-4518, Aug. 2024, doi: 10.1109/TMTT.2023.3349143.

[6] S. K. Vyas, S. Maurya, and V. P. Singh, "Electromagnetic and Particle-in-Cell Simulation Studies of a High Power Strap and Vane CW Magnetron," in *IEEE Plasma Sci*, vol. 42, no. 10, pp. 3373-3379, Oct. 2014, doi: 10.1109/TPS.2014.2352653.

[7] H. Huang, K. Huang and C. Liu, "Experimental Study on the Phase Deviation of 20-kW S-Band CW Phase-Locked Magnetrons," in *IEEE Microw. Wireless Compon. Lett.*, vol. 28, no. 6, pp. 509-511, Jun. 2018, doi: 10.1109/LMWC.2018.2832012.

[8] W. Li et al., "Particle-in-Cell Simulations of Injection Locking for S-Band Oven Magnetron Using Ultralow Energy," in *IEEE Plasma Sci*, vol. 53, no. 2, pp. 245-251, Feb. 2025, doi: 10.1109/TPS.2025.3532985.

[9] S. -T. Han, D. Kim, J. -S. Kim and J. -R. Yang, "Stabilization of Phase and Frequency of an S-Band Magnetron by Injection Locking," *2020 IEEE 21st International Conference on Vacuum Electronics (IVEC)*, Monterey, CA, USA, 2020, pp. 287-288, doi: 10.1109/IVEC45766.2020.9520523.

[10] Li, K., Zhang, Y., Zhu, H. C., Huang, K. M., and Yang, Y, (2019). "Theoretical and experimental study on frequency pushing effect of magnetron," in *Chin. Phys. B*, vol. 28, no. 11, pp: 118402, 2019, doi: 10.1088/1674-1056/ab4d3d

[11] Z. Liu et al., "Frequency-Locking Characteristics of Magnetron Locked by Injection Start-Up Technology," in *IEEE Trans. Electron Devices*, vol. 71, no. 11, pp. 7037-7042, Nov. 2024, doi: 10.1109/TED.2024.3453225.

[12] M.A. Shahid et al. "A phase‐shifterless experimental scheme for power combination of two magnetrons utilizing a single RF power amplifier." In *Microw. Opt. Technol. Lett.* vol. 64, no. 7, pp: 1192-1196, Apr. 2022, doi: 10.1002/mop.33269.

[13] C. Liu, H. Huang, Z. Liu, F. Huo and K. Huang, "Experimental Study on Microwave Power Combining Based on Injection-Locked 15-kW S-Band Continuous-Wave Magnetrons," in *IEEE Plasma Sci.*, vol. 44, no. 8, pp. 1291-1297, Aug. 2016, doi: 10.1109/TPS.2016.2565564.

[14] Z. Liu, X. Chen, M. Yang, P. Wu, K. Huang and C. Liu, "Experimental Studies on a Four-Way Microwave Power Combining System Based on Hybrid Injection-Locked 20-kW S-Band Magnetrons," in *IEEE Plasma Sci.*, vol. 47, no. 1, pp. 243-250, Jan. 2019, doi: 10.1109/TPS.2018.2876574

[15] H. Huang, B. Yang, N. Shinohara and C. Liu, "Coherent Power Combining of Four-Way Injection-Locked 5.8-GHz Magnetrons Based on a Five-Port Hybrid Waveguide Combiner," in *IEEE Trans. Microw. Theory Techn.*, vol. 72, no. 7, pp. 4395-4404, July 2024, doi: 10.1109/TMTT.2023.3347549.

[16] T. A. Treado et al., "Experimental results of power combining and phase-locking magnetrons for accelerator applications," in *Tech. Dig. - Int. Electron Devices Meet.*, San Francisco, CA, USA, 1990, pp. 541-544, doi: 10.1109/IEDM.1990.237139.

[17] Yuan, P., Zhang, Y., Ye, W., Zhu, H., Huang, K., and Yang, Y, "Power-combining based on master–slave injection-locking magnetron," in *Chin. Phys. B*, vol. 25, no. 7, pp: 078402, 2016, Jun. 2016, doi: 10.1088/1674-1056/25/7/078402

[18] X. Chen, B. Yang, N. Shinohara and C. Liu, "A High-Efficiency Microwave Power Combining System Based on Frequency-Tuning Injection-Locked Magnetrons," in *IEEE Trans. Electron Devices*, vol. 67, no. 10, pp. 4447-4452, Oct. 2020, doi: 10.1109/TED.2020.3013510.

[19] M. A. Shahid, T. M. Khan, M. Arif and M. Qamar-ul-Hassan, "A Power Combination System Based on Phase-Shifterless Scheme for Four Injection Locked S-Band Magnetrons," in *IEEE Plasma Sci.*, vol. 52, no. 3, pp. 744-752, March 2024, doi: 10.1109/TPS.2024.3366181.

[20] Y. Zhang, K. Huang, D. K. Agrawal, T. Slawecki, H. Zhu and Y. Yang, "Microwave Power System Based on a Combination of Two Magnetrons," in IEEE *Trans. Electron Devices*, vol. 64, no. 10, pp. 4272-4278, Oct. 2017, doi: 10.1109/TED.2017.2737555.

[21] P. Pengvanich, Y. Y. Lau; E. Cruz, R. M. Gilgenbach, B. Hoff; J. W. Luginslandl. "Analysis of peer-to-peer locking of magnetrons," in *Phys. Plasmas*, vol.15, no.10, Oct. 2008, doi: /10.1063/1.2992526.

[22] E. J. Cruz, B. W. Hoff, P. Pengvanich, Y. Y. Lau, R. M. Gilgenbach, J. W. Luginsland; "Experiments on peer-to-peer locking of magnetrons," in *Appl. Phys. Lett*, vol. 95, no. 19, Nov. 2009; doi: /10.1063/1.3262970.

[23] J. Liu et al., "Power Combining of Dual X-Band Coaxial Magnetrons Based on Peer-to-Peer Locking," in *IEEE Trans. Electron Devices*, vol. 68, no. 12, pp. 6518-6524, Dec. 2021, doi: 10.1109/TED.2021.3121225.

[24] R. Cheng et al., "Phase-Locked All-Cavity-Extraction Relativistic Magnetrons With Azimuth-Uniform Coupling Structure for Coherent Combining Applications," in *IEEE Trans. Electron Devices*, vol. 70, no. 11, pp. 5890-5896, Nov. 2023, doi: 10.1109/TED.2023.3309283.

[25] W. Li et al., "Phase Control Demonstration of S-Band Hybrid Phase-Locking Magnetrons for Array Applications," in *IEEE Trans. Microw. Theory Techn*, doi: 10.1109/TMTT.2025.3536474.

[26] Collins, George B. Microwave magnetrons. McGraw-Hill Book Co, 1948.

[27] J. C. Slater, "The phasing of magnetrons," Res. Lab. Electron., MIT, Cambridge, MA, USA, Tech. Rep. TK7855.M41 R43 no.35, April 1947.

[28] S. Wang, Y. Zhao, X. Chen and C. Liu, "A Novel Stability Improvement Method of S-Band Magnetron Systems Based on Its Anode Current Feature," in *IEEE Trans. Microw. Theory Techn*, vol. 72, no. 9, pp. 5530-5539, Sept. 2024, doi: 10.1109/TMTT.2024.3371160.

[29] Z. Liu et al., "Phase-Shifterless Power Controlled Combining Based on 20-kW S-Band Magnetrons With an Asymmetric Injection," in *IEEE Trans. Electron Devices*, vol. 39, no. 9, pp. 1425-1428, Sept. 2018, doi: 10.1109/LED.2018.2857808.

[30] X. Chen, B. Yang, N. Shinohara and C. Liu, "Low-Noise Dual-Way Magnetron Power-Combining System Using an Asymmetric H-Plane Tee and Closed-Loop Phase Compensation," in *IEEE Trans. Microw. Theory Techn*, vol. 69, no. 4, pp. 2267-2278, Apr. 2021, doi: 10.1109/TMTT.2021.3056550.

[31] C. Lai et al., "Highly Efficient Microwave Power System of Magnetrons Utilizing Frequency-Searching Injection-Locking Technique With No Phase Shifter," in *IEEE Trans. Microw. Theory Techn*, vol. 68, no. 10, pp. 4424-4432, Oct. 2020, doi: 10.1109/TMTT.2020.3006488.